\documentclass[a4paper,12pt]{article}
\usepackage{amssymb,amsmath,amsfonts,placeins,pbox,multirow,mathtools,bbold}
\usepackage{graphicx,rotate,color,slashed,cite,epstopdf,verbatim,url,multirow,booktabs}
\usepackage{epstopdf}
\usepackage{subfig}
\usepackage{graphicx,caption}
\usepackage{array}

\usepackage[colorlinks=true,
            linkcolor=magenta,
            urlcolor=blue,
            citecolor=blue]{hyperref}

\numberwithin{equation}{section}

\newcommand{\GHC}{$\mathrm{G}_{\rm HC}$}

\def\ssbh#1//#2//{\ensuremath{\xrightarrow [\substack{#2}]
    {\parbox{3cm}{\hfil $\scriptstyle \langle #1 \rangle$ \hfil}}}}

\let\CapTion=\caption
\def\caption#1{\CapTion{\em #1}}

\evensidemargin=\oddsidemargin
\title{\bf Vacuum misalignment in presence of four-Fermi operators}

\author{\bf Avik Banerjee\thanks{avik@chalmers.se}\,, Gabriele Ferretti\thanks{ferretti@chalmers.se}
\\ \medskip 
{}\small\em Department of Physics, Chalmers University of Technology, Fysikg\aa rden, 41296 G\"oteborg, Sweden}
\date{}

\begin{document}

\maketitle


\begin{abstract}
 We consider the issue of vacuum misalignment induced by four-Fermi couplings in a generic strongly coupled four-dimensional gauge theory. After briefly reviewing the general formalism, we focus on the case of partial compositness-like operators at leading order, which is relevant in applications to phenomenology.
We show that the interactions between an elementary fermion and composite spin-1/2 operators in various representations contribute to the effective potential with relative sign differences. Thus the correct sign required to misalign the vacuum is guaranteed to occur for some representations but not all of them. The overall sign dictating the specific representations responsible for misalignment can in principle be determined on the lattice. We also comment on the likely sign for some simple cases.
\end{abstract}


\bigskip

\newpage
\tableofcontents
\bigskip
\hrule
\bigskip

\section{Introduction}
\label{intro}
The dynamics of strongly coupled gauge theories is at the heart of many unsolved problems in theoretical physics. Even understanding the vacuum structure of such theories has proven challenging in general. Motivated by application to electroweak symmetry breaking (EWSB), we consider the issue of vacuum misalignment in composite Higgs-like theories. This phenomenon occurs when the theory is deformed by a weak perturbation invariant under a subgroup of the unbroken symmetries, but the vacuum of the strong dynamics does not allow to preserve the whole invariance and some weak generators are spontaneously broken. This is to be contrasted with Technicolor~\cite{Dimopoulos:1979es,Susskind:1978ms}, where the symmetry of the weak sector cannot be fully embedded into the  unbroken subgroup and thus some of its generators are guaranteed to be broken. In this work we refer to vacuum misalignment only in the former sense.

Given what we now know about the Higgs sector, this type of vacuum misalignment is a plausible way to achieve EWSB via a strongly coupled theory. It was proposed by Kaplan and Georgi in~\cite{Kaplan:1983fs}, although their original realization still involves elementary scalars, so it is less appealing as a solution to the hierarchy problem. It was quickly pointed out, e.g. in~\cite{Banks:1984gj} that in scalarless theories, gauge fields are not enough to achieve misalignment. This fact is closely related to the impossibility of breaking a vector-like symmetry in vector-like theories~\cite{Vafa:1983tf,Witten:1983ut,Weingarten:1983uj}. A way around this problem was suggested in~\cite{Dugan:1984hq} where EWSB  was achieved with only gauge fields by gauging an additional (chiral) group.

The solution of~\cite{Dugan:1984hq} comes at the cost of introducing gauge anomalies that need to be canceled by additional spectator fermions. Moreover, it does not allow to have non-zero hyperquark masses, and it predicts additional heavy vector bosons that are now severely constrained by direct searches at LHC.
Finally, in~\cite{ArkaniHamed:2002qy,Contino:2003ve,Agashe:2004rs} it was shown that one could use couplings between the Standard Model fermions and composite fermions of the strong sector to achieve the goal of misalignment, a proposal inspired by Kaplan's idea of partial compositeness~\cite{Kaplan:1991dc}. Although this was initially framed in the context of extra-dimensional theories, purely four-dimensional gauge theories that have a chance of realizing both mechanisms~\cite{Kaplan:1983fs,Kaplan:1991dc} were constructed more recently in~\cite{Barnard:2013zea,Ferretti:2013kya} and following works.

In this work we ask the question of what can be said in general about vacuum misalignment induced by fermionic interactions in the case of four-dimensional gauge theories. The main hurdle is the difficulty to estimate the sign of the contribution to the pseudo-Nambu--Goldstone boson (pNGB) masses. This sign cannot be determined in the effective theory and requires a study of the underlying dynamics. So far, the only result we are aware of is that of Golterman and Shamir~\cite{Golterman:2015zwa} showing that misalignment may occur (for sufficiently large coupling and anomalous dimensions) in a $SU(5)/SO(5)$ global symmetry breaking coset~\cite{Ferretti:2014qta} with composite fermions in the $\mathbf{5}$ representation at NLO. (Since the leading contribution vanishes by symmetry arguments.)

Here we would like to discuss the cases where the LO contribution to the pNGB potential does not vanish. We are facing the problem that the overall sign to the pNGB masses is determined by the strong dynamics and therefore hard to pin down. However we show that composite fermions in different representations contribute to the masses with different relative signs, purely at the group theory level. This means that, regardless of the overall sign of the effective Hamiltonian, there will always be some representation that misaligns the vacuum. A further dynamical assumption, that we find reasonable, namely that singlet fermions should not misalign the vacuum, allows to speculate some positivity constraints on the overall coefficients, leading to conjectural inequalities between spectral functions in various channels.

There are plenty of models with only one irrep of hyperfermions where some of these issues could be investigated on the lattice, the main one being $Sp(2N)$ hypercolor with hyperfermions in the anti-symmetric.  Of course, the actual occurrence of misalignment in a specific model still hinges on the dynamical issue of whether the dimension six operators responsible for it are sufficiently enhanced in the infra-red, an issue that has so far only been addressed perturbatively~\cite{DeGrand:2015yna,BuarqueFranzosi:2019eee}.

In fact, existing lattice results for phenomenologically interesting scenarios with two irreps of hyperfermions can be of help to estimate the overall sign of the pNGB masses. For $SU(4)$ \cite{Ayyar:2018zuk} and $Sp(4)$ \cite{Hsiao:2022kxf} hypercolor theories, the lattice calculations of the spectrum of chimera baryons can provide a heuristic understanding of which representations of the fermions may misalign the vacuum in composite Higgs models.

The paper is organized as follows. After briefly reviewing the general formalism in Sect.~\ref{general}, we discuss the case of interest in Sect.~\ref{partialcompo} and provide illustrative examples in Sect.~\ref{examples}. We conclude with some more general comments and an outlook for future developments.

\section{General Formalism}
\label{general}
In this section we briefly review some well-known facts about the dynamics of strongly coupled gauge theories in order to set up the notation.
The theories of interest are four-dimensional gauge theories with a simple ``hypercolor'' gauge group \GHC~and a set of left-handed Weyl fermions $\Psi$ in a non-chiral (self-conjugate) representation of \GHC. These fermions may thus belong to a real (R) or pseudoreal (PR) representation, in which case we write $\Psi^i$ where the flavor index $i$ runs from $1$ to $N$ for the R or $2N$ for the PR. Alternatively, they may belong to a complex (C) representation $\Psi^i=\psi^i$ and its conjugate $\Psi^{(i+N)}=\tilde\psi_i$, where $i=1\dots N$. 

In the UV, the hypercolor invariant Lagrangian can be written as 
\begin{align}
    \mathcal{L}_{\rm UV} & = i\Psi^\dagger_i \bar{\sigma}^\mu D_\mu \Psi^i - \left(m_{ij}\Psi^i\Psi^j +  m^{\dagger ij}\Psi^\dagger_i\Psi^\dagger_j\right), \label{Lagrangian}
\end{align}
where $m$ is a \GHC~invariant mass term for the hyperfermions that can always be brought to the form
\begin{align}
m=\left\{ \begin{array}{lll}
		\frac{1}{2}\,\mathrm{diag}(m_1, \dots, m_N) &  & (R),
		\\
		\frac{1}{2}\,\mathrm{diag}(m_1, \dots, m_N)\otimes i\sigma^2 & &  (PR),
            \\ 
            \sigma^1\otimes \mathrm{diag}(m_1, \dots, m_N)  & & (C),
	\end{array} \right.
        \label{canonicalmass}
\end{align}
where $m_i\geq 0$, for all $i=1\dots N$. In this work we consider only CP preserving theories, corresponding to choosing a hypercolor theta angle $\theta=0$ with the above choice of masses.

There may be other fermions in the theory, but we focus on the symmetry breaking pattern associated to the $\Psi$s.
In the case where all $\Psi$ are massless it is well known that, in the appropriate dynamical region, the global symmetry breaking patterns associated to the three choices of representation are $SU(N)/SO(N)$, $SU(2N)/Sp(2N)$ and $SU(N)\times SU(N)/SU(N)$ respectively. 
In all three cases we are dealing with a symmetric coset $G/H$ where the generators $T^A$ of $G$ can be divided into the broken  $T^{\hat{a}}$ and unbroken $T^a$ generators satisfying the algebra 
\begin{align}
    [T^a,T^b]=if^{abc}T^c\,,
    \quad
    [T^{\hat{a}},T^{\hat{b}}]=if^{\hat{a}\hat{b}c}T^c\,,
    \quad
    [T^a,T^{\hat{b}}]&=if^{a\hat{b}\hat{c}}T^{\hat{c}}.\label{symmstriccoset}
\end{align}

The currents and corresponding charges (operators in the Fock space of the QFT) can be written as
\begin{align}
    J^A_{\mu} = \Psi^\dagger_i\bar{\sigma}_\mu(T^{A})^i_j\Psi^j, \qquad Q^A =\int d^3x~ J^A_{0}.
\end{align}
The charges obey exactly the same commutation relations as (\ref{symmstriccoset}). Let the vacuum state annihilated by the $Q^a$ be denoted by $|{\mathrm{vac}}\rangle_0$. The first well known fact is that the explicit embedding of $H$ into $G$ is arbitrary and one could use any isomorphic subgroup $gHg^\dagger$, where $g\in G$. Equivalently, one could rotate the vacuum state to 
\begin{equation}
|{\mathrm{vac}}\rangle_\Pi = \exp{(i \Pi^{\hat{a}} Q^{\hat{a}})}|{\mathrm{vac}}\rangle_0,
\end{equation}
and conjugate all the operators accordingly. For convenience we use dimensionless fields $\Pi^{\hat{a}}=\pi^{\hat{a}}/f$, where $\pi$ are the pseudo-Nambu--Goldstone bosons (pNGBs) and $f$ their decay constant.

The invariant tensor $I_0^{ij}$ associated with the $H$ invariant vacuum $|{\mathrm{vac}}\rangle_0$ is defined through
\begin{align}
 _0\langle{\mathrm{vac}}|\Psi^i \Psi^j|{\mathrm{vac}}\rangle_0 = - B I_0^{ij},
\end{align}
with
\begin{align}
I_0 = \mathbf{1}\, (R),\qquad
\mathbf{1}\otimes i\sigma^2\, (PR), \qquad \sigma^1\otimes \mathbf{1}\, (C) \label{canonicalvac}
\end{align}
for the three cases and $B$ a low energy constant of dimension $[M]^3$. Having chosen this specific expression for $I_0$, the broken/unbroken generators are identified by the relations $T^{\hat{a}} I_0 - I_0 T^{{\hat{a}}T}=0$ and $T^a I_0 + I_0 T^{aT}=0$ respectively.

If we were to perturb the theory by a fully $G$ invariant Hamiltonian $\mathcal{H}$, obeying $[Q^A,\mathcal{H}]=0$, no potential for the $\Pi$s would be generated since
$_\Pi\langle{\mathrm{vac}}|\mathcal{H}|{\mathrm{vac}}\rangle_\Pi=\;_0\langle{\mathrm{vac}}|\mathcal{H}|{\mathrm{vac}}\rangle_0={\mathrm{const.}}$
If, on the other hand, the Hamiltonian is invariant under $H$ but does not fully commute with the broken generators, by expanding the exponentials one obtains~\cite{Dashen:1970et,Weinberg:1975gm}
\begin{align}
V(\Pi)=\;_\Pi\langle{\mathrm{vac}}|\mathcal{H}|{\mathrm{vac}}\rangle_\Pi=&\,_0\langle{\mathrm{vac}}|\mathcal{H}|{\mathrm{vac}}\rangle_0 - i \Pi^{\hat{a}}\; _0\langle{\mathrm{vac}}|[Q^{\hat{a}},\mathcal{H}]|{\mathrm{vac}}\rangle_0\nonumber\\& -\frac{1}{2}\Pi^{\hat{a}} \Pi^{\hat{b}}\;  _0\langle{\mathrm{vac}}|[Q^{\hat{a}},[Q^{\hat{b}},\mathcal{H}]]|{\mathrm{vac}}\rangle_0 +\dots\label{dashen}
\end{align}
In order for $|{\mathrm{vac}}\rangle_0$ to be an acceptable vacuum we need the potential to be minimized at $\Pi=0$, i.e.
\begin{align}
&_0\langle{\mathrm{vac}}|[Q^{\hat{a}},\mathcal{H}]|{\mathrm{vac}}\rangle_0 = 0, \qquad & &\mbox{(``no-tadpole condition'')}\label{notadpole}\\
&(M^2)^{\hat{a}\hat{b}}=\; -\frac{1}{f^2}\;_0\langle{\mathrm{vac}}|[Q^{\hat{a}},[Q^{\hat{b}},\mathcal{H}]]|{\mathrm{vac}}\rangle_0 \geq 0\qquad & &\mbox{(``no-tachyon condition'')} \label{notachyon}.
\end{align}

The mass term in \eqref{Lagrangian} can also be interpreted as a perturbation Hamiltonian ${\mathcal{H}}_{\mathrm{mass}}= m_{ij}\Psi^i\Psi^j +  m^{\dagger ij}\Psi^\dagger_i\Psi^\dagger_j$, breaking $H$ to the subalgebra generated by $T^\mathrm{a}$ leaving $m$ invariant. In other words $[Q^{\mathrm{a}}, {\mathcal{H}}_{\mathrm{mass}}]=0 \Rightarrow T^{\mathrm{a}} m + m T^{\mathrm{a}T}=0$, where $T^\mathrm{a}$ spans a subset of the unbroken generators $T^a$. The choices \eqref{canonicalmass} and \eqref{canonicalvac} are compatible with the absence of tadpoles and tachyons (\ref{notadpole}) and (\ref{notachyon}), so no pNGB acquires a vacuum expectation value and their masses are given by the Gell-Mann--Oakes--Renner formula~\cite{Gell-Mann:1968hlm}. 

If one choose instead to perturb by a generic complex symmetric mass $m_{ij}$, the vacuum (\ref{canonicalvac}) would no longer be suitable, since it would fail some of the criteria  (\ref{notadpole}), and  (\ref{notachyon}). However it would always be possible to redefine $I_0$ and the generators $T^A$ by a global $G$ transformation and, possibly, by an additional anomalous $U(1)$: $\Psi^i\to e^{i\alpha}\Psi^i$.

\begin{figure}[t]
\begin{center}
            \includegraphics[trim=60 550 60 75,clip,width=0.6\linewidth]{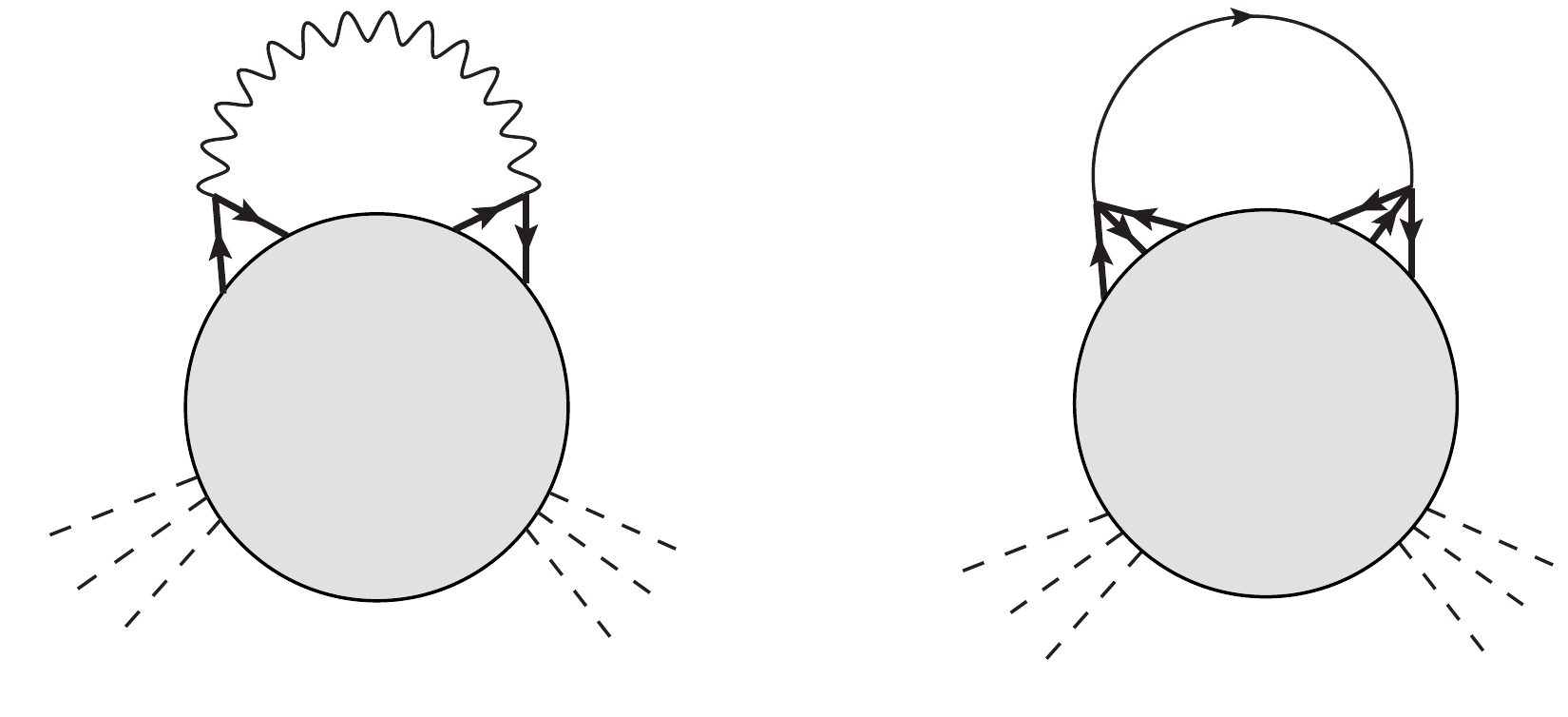}
\end{center}    	
    \caption{\sf\it Leading order diagrams contributing to the pNGB potential. Left: Contribution from the gauge interactions; Right: contribution from four-Fermi interactions. Similar diagrams are referred to as ``cat graphs'' in  \cite{Kaplan:1983fs} although we would like to suggest the more accurate description of  \href{https://tenor.com/view/cosmos-cat-snuggles-meow-meditation-gif-14408304}{``Cosmic cat diagrams''}. Thick lines represent hyperfermions and each pNGB insertion, corresponding to a commutator with the charges, is represented by a whisker.}
    \label{fig:cat}
\end{figure}

Another important source of perturbation comes from the weak gauging of a subgroup $G_{\rm w}$ of $H$. The hyperfermion mass term should at least preserve the gauge symmetry, implying that the gauged generators are also given by $T^{\mathrm{a}}$ or a subset of them. The perturbed Hamiltonian arising from the gauge contribution, pictorially depicted in the left panel of figure \ref{fig:cat}, can be written as~\cite{Das:1967it,Weinberg:1975gm,Peskin:1980gc,Preskill:1980mz}
\begin{align}
    \mathcal{H}_{\rm gauge} &=-\frac{i}{2}\mathcal{G}_{A\mathrm{a}} \mathcal{G}_{B\mathrm{b}}\int d^4x ~\Delta^{\mu\nu}(x) \delta^{\mathrm{a b}}  \, T \left\{J^A_\mu(x)J^B_\nu(0)\right\},
\label{gauge_Hamiltonian}
\end{align}
where $T\{...\}$ denotes time ordering and $\Delta^{\mu\nu}(x)\delta^{\mathrm{a b}} $ is the propagator of the massless weak gauge fields $A^{\mathrm{a}}_\mu$. The couplings $\mathcal{G}_{A\mathrm{a}}$ transform in the adjoint representations of $G$ and $G_{\rm w}$ and their physical values are given by $\mathcal{G}_{A\mathrm{a}}= g \delta_{A\mathrm{a}}$. 

The joint expectation values of the product of two current can be written as
\begin{align}
    _0\langle{\mathrm{vac}}|T\left\{J^{a}_\mu(x)J^{b}_\nu(0)\right\}|{\mathrm{vac}}\rangle_0 & = G_{\mu\nu}(x) \delta^{ab} \\
    _0\langle{\mathrm{vac}}|T\left\{J^{\hat{a}}_\mu(x)J^{\hat{b}}_\nu(0)\right\}|{\mathrm{vac}}\rangle_0 & = \hat{G}_{\mu\nu}(x) \delta^{\hat{a}\hat{b}} \\
    _0\langle{\mathrm{vac}}|T\left\{J^{\hat{a}}_\mu(x)J^{b}_\nu(0)\right\}|{\mathrm{vac}}\rangle_0 & = 0, \label{JaJahat}
\end{align}
where the two-point functions $G_{\mu\nu}(x)$ and $\hat{G}_{\mu\nu}(x)$ can be expressed in terms of their spectral representations \cite{Weinberg:1967kj}. The no-tadpole condition  (\ref{notadpole}) can be realized by noticing that the joint expectation value of a broken and unbroken current is zero (\ref{JaJahat}).
The pNGB mass matrix arising from $\mathcal{H}_{\rm gauge}$ is given by~\cite{Das:1967it,Weinberg:1975gm,Peskin:1980gc,Preskill:1980mz}
\begin{align}
   (M^2)^{\hat{a}\hat{b}}=\; i\frac{g^2}{f^2}f^{\hat{a}\hat{c}\mathrm{a}}f^{\hat{b}\hat{c}\mathrm{a}}\int d^4x ~\Delta^{\mu\nu}(x) \, \left[G_{\mu\nu}(x)-\hat{G}_{\mu\nu}(x)\right]\,,
   \label{pNGB_mass_gauge}
\end{align}
which has positive eigenvalues only, as shown in \cite{Witten:1983ut}. Thus, the vacuum $|{\mathrm{vac}}\rangle_0$ is not destabilized by weakly gauging $G_{\rm w}$. This is a positive feature in the case of QCD, since one does not want it to break electromagnetic interactions.

\section{Vacuum misalignment via partial compositeness }
\label{partialcompo}

Contrary to QCD, in applications to composite Higgs models~\cite{Kaplan:1983fs} a dynamical mechanism to ``misalign'' the vacuum and trigger EWSB is necessary. As discussed in the previous section, the mass terms or the gauge contributions can not misalign the vacuum in a vector-like gauge theory. The interactions of \GHC~singlet elementary fermions provide another source of explicit breaking of $G$. In this section we show that four-Fermi interactions between composite fermions of the strong sector and elementary chiral fermions can in fact misalign the vacuum. There is a close analogy between the currents and the gauge fields of the previous section and the composite and elementary fermions considered here. 

The composite fermions are hypercolor singlet bound states of three hyperfermions (schematically $\Psi\Psi\Psi$) which may transform as a one-, two- or three-index representations of $G$. Note that the bound state may also involve hyperfermions in other irreps of \GHC~as in the case of top-partners in the composite Higgs models. For the purpose of our analysis, the exact composition of the bound state fermions is not relevant, only the $G$-irrep under which they transform is important. We denote the \GHC~singlet composite fermions by $B^R_\alpha,\, B^\dagger_{R\dot\alpha}$ where the index $R$ spans a specific irrep of $G$ which we also denote by $R$ for brevity. Weyl indices are denoted as $\alpha$ and $\dot\alpha$.

Commutation relations between $B^R_\alpha$  and the charges can be written as
\begin{align}
    [Q^A,B^R_\alpha]=-(T^A)^R_S B^S_\alpha\,, \qquad [Q^A,B^\dagger_{\dot{\alpha}R}]=B^\dagger_{\dot{\alpha}S} (T^A)^S_R\,.
    \label{QB_comm}
\end{align}
In the above equation, the generators $T^A$ (with $A=1, \dots \mathrm{dim}(G)$) are in the $\mathrm{dim}(R)$ dimensional representation of $G$.

In addition to the composite states, we consider elementary chiral fermions $\lambda^{\mathrm{i}\alpha}$ charged under an irrep `$\mathrm{i}$' of a subgroup of $H$. A particularly interesting example is when the fermions are charged under the weakly gauged subgroup $G_{\rm w}$, in which case one might have to introduce additional spectator fermions to cancel gauge anomalies.  
Though we discuss this specific scenario, our results are valid even in the absence of gauging.

The linear mixing of elementary fermions with their composite counterpart through four-Fermi interactions ($\lambda\Psi\Psi\Psi$), referred to as partial compositeness~\cite{Kaplan:1991dc}, generically break the global symmetry $G$ of the strong sector.
The low energy Lagrangian describing partial compositeness is given by
\begin{align}
    \mathcal{L}_{\rm PC}= y \mathcal{P}^{\dagger}_{R\mathrm{i}} \lambda^{\mathrm{i}\alpha} B^R_\alpha + {\rm h.c.} \label{pc}
\end{align}
where $\mathcal{P}_{R\mathrm{i}}$ is a projector matrix and $y$ is a real dimensionful interaction strength. 
For simplicity we consider a single coupling $y$, for a more general case see~\cite{Golterman:2017vdj}.

The effective pNGB potential generated from \eqref{pc} at LO can be constructed using the operators in the last column of table \ref{decomp}. However, the signs of the coefficients associated with these operators, which are determined by the strong dynamics, can not be fixed using this effective construction. Our main objective is to estimate these signs in a manner similar to the gauge contribution discussed in the previous section.   

The perturbed Hamiltonian arising from (\ref{pc}) as shown in the right panel of figure \ref{fig:cat} is
\begin{align}
    \mathcal{H}_{\rm PC} & =-\frac{i}{2}y^2\int d^4x \Delta^{\dot{\alpha}\alpha}(x) \delta^{\mathrm{i}}_{\mathrm{j}} \, T\left\{\mathcal{P}^\dagger_{R\mathrm{i}} B^R_\alpha(x) B^\dagger_{Q\dot{\alpha}}(0) \mathcal{P}^{Q\mathrm{j}} + {\rm h.c.} \right\},
\end{align}
where $\Delta^{\dot{\alpha}\alpha}(x) \delta^{\mathrm{i}}_{\mathrm{j}}$ is the $\lambda$ propagator.

\begin{table}
\centering
    \begin{tabular}{ccccc}
    \hline\hline
    $SU(N)$ & $\to$ & $SO(N)$ & \\
    \hline 
    $\mathbf{Ad}$ & & $\mathbf{Ad}+\mathbf{S}_2$ & ${\rm tr}(\mathcal{P}U\mathcal{P}^*U^*)$ \\
    $\mathbf{S}_2$ & & $\mathbf{1}+\mathbf{S}_2$ & ${\rm tr}(\mathcal{P}U^*){\rm tr}(\mathcal{P}^*U)$ \\
    \hline\hline
    $SU(2N)$ & $\to$ & $Sp(2N)$\\
    \hline 
    $\mathbf{Ad}$ & & $\mathbf{Ad}+\mathbf{A}_2$ & ${\rm tr}(\mathcal{P}U\mathcal{P}^*U^*)$\\
    $\mathbf{A}_2$ & & $\mathbf{1}+\mathbf{A}_2$ & ${\rm tr}(\mathcal{P}U^*){\rm tr}(\mathcal{P}^*U)$\\
    \hline\hline
     $SU(N)\times SU(N)$ & $\to$ & $SU(N)$\\
    \hline 
    $(\mathbf{F}, \mathbf{F}) $ & & $\mathbf{A}_2+\mathbf{S}_2$ & ${\rm tr}(U\mathcal{P}^TU^*\mathcal{P}^\dagger)$\\
    $(\mathbf{F}, \overline{\mathbf{F}}) $ & & $\mathbf{1}+\mathbf{Ad}$ & ${\rm tr}(\mathcal{P}U^\dagger){\rm tr}(\mathcal{P}^\dagger U)$ \\
    \hline\hline
    \end{tabular}
    \caption{\sf\it List of the $G$-irreps up to two indices that split in two $H$-irreps. The symbols $\mathbf{1},\mathbf{S}_2,\mathbf{A}_2,\mathbf{Ad}, \mathbf{F}, \overline{\mathbf{F}}$ stand for the singlet, symmetric, anti-symmetric, adjoint, fundamental, and antifundamental of the respective groups. The last column shows the LO operators contributing to the effective pNGB potential, where $U$ is the unitary matrix denoting the nonlinear realization of the pNGBs around $\Pi=0$. }
    \label{decomp}
\end{table}
While the composite fermion operators transform under $R$ of $G$, the vacuum breaks spontaneously $G$ to $H$. Thus, the vacuum correlation functions $\langle B B^\dagger \rangle$ are classified by $H$-irreps and we must decompose $R$ into $H$-irreps. We consider the decomposition $R\to r+r^\prime$, which captures all the relevant cases for the three types of cosets up to two-index representations of $R$ as shown in table~\ref{decomp}.

Using this notation, the $\langle B B^\dagger \rangle$ correlators are given by
\begin{align}
    _0\langle{\mathrm{vac}}|T\left\{B^r_\alpha(x) B^\dagger_{q\dot{\alpha}}(0)\right\}|{\mathrm{vac}}\rangle_0 & = G_{\alpha\dot{\alpha}}(x) \delta^{r}_q \\
    _0\langle{\mathrm{vac}}|T\left\{B^{r^\prime}_\alpha(x) B^\dagger_{{q^\prime}\dot{\alpha}}(0)\right\}|{\mathrm{vac}}\rangle_0 & = G^{\prime}_{\alpha\dot{\alpha}}(x) \delta^{r^\prime}_{q^\prime} \\
    _0\langle{\mathrm{vac}}|T\left\{B^{r}_\alpha(x) B^\dagger_{q^\prime\dot{\alpha}}(0)\right\}|{\mathrm{vac}}\rangle_0 & = 0.
\end{align}
Similarly to the currents, the two point functions $G_{\alpha\dot{\alpha}}(x)$ and $G^\prime_{\alpha\dot{\alpha}}(x)$ can be expressed in terms of their spectral functions $\rho_G(\mu^2)$ and $\rho_{G^\prime}(\mu^2)$. The Fourier transform of $G_{\alpha\Dot{\alpha}}(x)$ is given by
\begin{align}
    \int d^4x \, e^{ikx} \, G_{\alpha\Dot{\alpha}}(x) = i k^\mu (\sigma_\mu)_{\alpha\Dot{\alpha}} \int d\mu^2 \frac{\rho_G(\mu^2)}{k^2-\mu^2+i\epsilon}\,,
\end{align}
and similarly for $G^\prime_{\alpha\dot{\alpha}}(x)$. In the short distance limit, restoration of the global symmetry $G$ implies for the correlators $G_{\alpha\Dot{\alpha}}(x) - G^\prime_{\alpha\Dot{\alpha}}(x) \to 0$. If we assume that the composite fermion operators behave as free fermion fields in this limit ($k^2\to \infty$) a sum rule for the spectral functions, analogous to the Weinberg's sum rules \cite{Weinberg:1967kj, PhysRevLett.18.761, Bernard:1975cd},  can be obtained as
\begin{align}
    \int d\mu^2 \left[\rho_G(\mu^2)-\rho_{G^\prime}(\mu^2)\right] =0.
    \label{sumrule}
\end{align}

The remaining task is to check the conditions (\ref{notadpole}) and (\ref{notachyon}). We start with the tadpole equation which implies
\begin{align}
    \mathcal{P}^\dagger_{r\mathrm{i}} (T^{\hat{a}})^r_{q^\prime}\mathcal{P}^{q^\prime\mathrm{i}} - \mathcal{P}^\dagger_{r^\prime\mathrm{i}} (T^{\hat{a}})_q^{r^\prime}\mathcal{P}^{q\mathrm{i}} =0.
\label{PC_notadpole}
\end{align}
Similarly, the pNGB mass matrix at the $H$-invariant vacuum $|{\mathrm{vac}}\rangle_0$ is given by
\begin{align}
\nonumber
    (M^2)^{\hat{a}\hat{b}} &= \frac{y^2}{f^2}\mathcal{K}
    \left[ \mathcal{P}^\dagger_{r\mathrm{i}}  \left\{ (T^{\hat{b}})^r_{q^\prime} (T^{\hat{a}})^{q^\prime}_s + (T^{\hat{a}})^r_{q^\prime} (T^{\hat{b}})^{q^\prime}_s \right\} \mathcal{P}^{s\mathrm{i}} - \mathcal{P}^\dagger_{r^\prime\mathrm{i}}  \left\{ (T^{\hat{b}})^{r^\prime}_{q} (T^{\hat{a}})^{q}_{s^\prime} + (T^{\hat{a}})^{r^\prime}_{q} (T^{\hat{b}})^{q}_{s^\prime} 
 \right\} \mathcal{P}^{s^\prime\mathrm{i}} \right.\\
    & - \mathcal{P}^\dagger_{r\mathrm{i}}  \left\{ (T^{\hat{b}})^{r}_{q} (T^{\hat{a}})^{q}_{s^\prime} - (T^{\hat{a}})^{r}_{q^\prime} (T^{\hat{b}})^{q^\prime}_{s^\prime}  \right\} \mathcal{P}^{s^\prime\mathrm{i}} + \left.  \mathcal{P}^\dagger_{r^\prime\mathrm{i}} \left\{ (T^{\hat{b}})^{r^\prime}_{q^\prime} (T^{\hat{a}})^{q^\prime}_{s} - (T^{\hat{a}})^{r^\prime}_{q} (T^{\hat{b}})^{q}_{s}  \right\} \mathcal{P}^{s\mathrm{i}}\right].
    \label{PC_notachyon}
\end{align}
The overall coefficient $\mathcal{K}$ can be expressed in terms of the spectral functions $\rho_G(\mu^2)$ and $\rho_{G^\prime}(\mu^2)$ as
\begin{align}
    \mathcal{K}\equiv i\int d^4x ~\Delta^{\dot{\alpha}\alpha}(x) \, \left[ G_{\alpha\Dot{\alpha}}(x) -G^\prime_{\alpha\Dot{\alpha}}(x) \right] = 2\int \frac{d^4k_E}{(2\pi)^4} \int d\mu^2 \, \frac{\left[ \rho_G(\mu^2) -\rho_{G^\prime}(\mu^2)  \right]}{k^2_E+\mu^2},
    \label{spectral}
\end{align}
where $k_E$ denotes Euclidean momentum. Few comments regarding \eqref{PC_notachyon} and \eqref{spectral} are necessary. First of all, the second line of \eqref{PC_notachyon} vanishes if one chooses to couple $\lambda^{\mathrm{i}\alpha}$ only with $B^r_\alpha$ or $B^{r^\prime}_\alpha$ which implies either $\mathcal{P}^{r\mathrm{i}}$ or $\mathcal{P}^{r^\prime\mathrm{i}}$ is zero. This choice manifestly satisfies \eqref{PC_notadpole} as well. The remaining positive definite terms in the first line of \eqref{PC_notachyon} appear with a relative sign difference between them. 
Thus, irrespective of the sign of $\mathcal{K}$, we conclude that \eqref{PC_notachyon} violates the no-tachyon condition for one of the two $H$-irreps. Which of the two irreps misaligns the vacuum depends on the sign of $\mathcal{K}$ which can in principle be determined using lattice gauge theory.

Note that the gauge contribution \eqref{pNGB_mass_gauge} is a special case of the more generic \eqref{PC_notachyon} with appropriate modifications of $\mathcal{K}$. The weak gauge bosons transform under the adjoint of $G_{\rm w}$ which forces one to couple them with the unbroken currents in the adjoint of $H$, leading to a unique choice of the projector ($\mathcal{G}_{A\mathrm{a}}=g\delta_{A\mathrm{a}}$ in the notation of previous section). Thus, in contrast to the partial compositeness, gauge contributions yield a single term which is shown to be positive~\cite{Witten:1983ut}, satisfying the no-tadpole condition.   

To illustrate further, consider the $SU(N)/SO(N)$ coset and assume that $B^R_\alpha$ transforms under the symmetric $\mathbf{S}_2$ of $SU(N)$ which decomposes into a $\mathbf{1}$ ($B^{\mathbf{1}}_\alpha$) and a $\mathbf{S}_2$ ($B^{\mathbf{S}_2}_\alpha$) of $SO(N)$ (see table \ref{decomp}). The expression for the pNGB mass matrix is given by
\begin{align}
\nonumber
    (M^2)^{\hat{a}\hat{b}} &= \frac{8y^2}{Nf^2}\mathcal{K}
    \left[ {\rm tr}\left(\mathcal{P}^\dagger_{\mathrm{i}} T^{\hat{a}} T^{\hat{b}}\right) {\rm tr}\left(\mathcal{P}^{\mathrm{i}}\right) +{\rm tr}\left(\mathcal{P}^\dagger_{\mathrm{i}}\right){\rm tr}\left(T^{\hat{b}} T^{\hat{a}} \mathcal{P}^{\mathrm{i}} \right) \right.\\
    & \left. - {\rm tr}\left(\mathcal{P}^\dagger_{\mathrm{i}} T^{\hat{a}}\right){\rm tr}\left(T^{\hat{b}} \mathcal{P}^{\mathrm{i}} \right) - {\rm tr}\left(\mathcal{P}^\dagger_{\mathrm{i}} T^{\hat{b}}\right){\rm tr}\left(T^{\hat{a}} \mathcal{P}^{\mathrm{i}} \right) \right],
    \label{PC_mass_real}
\end{align}
where now we express both $\mathcal{P}$ and the generators $T^{\hat{a}}$ as $N\times N$ matrices and the trace is over the $SU(N)$ indices. The expression for $\mathcal{K}$ is given by \eqref{spectral} with $\rho_G(\mu^2)\equiv \rho_{\mathbf{1}}(\mu^2)$ and $\rho_{G^\prime}(\mu^2)\equiv \rho_{\mathbf{S}_2}(\mu^2)$ being the spectral functions associated with $B^{\mathbf{1}}_\alpha$ and $B^{\mathbf{S}_2}_\alpha$, respectively.

The choice of the projector matrix depends on how $\lambda^{\mathrm{i}\alpha}$ transforms under $G_{\rm w}\subset SO(N)$. If $\lambda^{\alpha}$ is a singlet of $G_{\rm w}$, $\mathcal{P}$ can be considered to be a diagonal matrix, $\mathcal{P}_{\mathbf{1}}=\mathbf{1}_N/\sqrt{N}$. This choice corresponds to coupling $\lambda^{\alpha}$ with $B^{\mathbf{1}}_\alpha$. In this case, the last two terms within square brackets of \eqref{PC_mass_real} vanish due to the traceless nature of the $SU(N)$ generators, while the first two terms are positive definite. In contrast, if $\lambda^{\mathrm{i}\alpha}$ transforms non-trivially under $G_{\rm w}$ one can choose $\mathcal{P}_{\mathbf{S}_2}$ as a $N\times N$ traceless symmetric matrix such that $\lambda^{\mathrm{i}\alpha}$ couples to $B^{\mathbf{S}_2}_\alpha$. For this choice, the first two terms within square brackets of \eqref{PC_mass_real} are zero. The presence of an overall sign difference with respect to the previous case indicates the possibility of tachyonic mass eigenvalues and therefore misalignment of the vacuum $|{\mathrm{vac}}\rangle_0$. In passing we note that the effective operator ${\rm tr}(\mathcal{P}U^*){\rm tr}(\mathcal{P}^*U)$ (see table \ref{decomp}) expanded around $\Pi=0$ leads to the same mass matrix as in \eqref{PC_mass_real}, however, the relative sign difference between the $\mathcal{P}_{\mathbf{1}}$ and $\mathcal{P}_{\mathbf{S}_2}$ contributions can not be predicted using the effective formalism. 

Although the sign of $\mathcal{K}$ is not directly calculable from \eqref{spectral}, in this specific case we could argue that it is a positive quantity. We do not expect the linear mixing between a $H$-singlet $B^{\mathbf{1}}_\alpha$ and a $G_{\rm w}$-singlet  $\lambda^{\alpha}$ to misalign the vacuum. This leads us to expect the following inequality involving the spectral functions $\rho_{\mathbf{1}}(\mu^2)$ and $\rho_{\mathbf{S}_2}(\mu^2)$
\begin{align}
    \int \frac{d^4k_E}{(2\pi)^4} \int d\mu^2 \, \frac{\left[ \rho_{\mathbf{1}}(\mu^2) -\rho_{\mathbf{S}_2}(\mu^2)  \right]}{k^2_E+\mu^2} >0.
    \label{spectral_cond}
\end{align}
Assuming the dominance of the leading resonance and considering the sum rule \eqref{sumrule}, one can approximate the spectral functions as $\rho_{\mathbf{1}}(\mu^2)\simeq C\delta(\mu^2-M_{\mathbf{1}}^2)$ and $\rho_{\mathbf{S}_2}(\mu^2)\simeq C\delta(\mu^2-M_{\mathbf{S}_2}^2)$, where $C$ is a positive constant. This ansatz for the spectral functions soften the UV divergence of $\mathcal{K}$ from quadratic to logarithmic. Further, the condition \eqref{spectral_cond} implies that $M_{\mathbf{S}_2}>M_{\mathbf{1}}$.

It is worth mentioning that couplings of the elementary fermion with a $B^{\mathbf{F}}_\alpha$ or $B^{\mathbf{A}_2}_\alpha$ of $G$ do not generate a potential for the pNGBs at the leading order. If instead, we choose to couple $\lambda^{\mathrm{i}\alpha}$ with $B^{\mathbf{Ad}}_\alpha$ of $G$, a possibility of vacuum misalignment with either $\mathbf{S}_2$ or $\mathbf{A}_2$ of $H$ will arise. In this case, however, it is less clear what to expect the sign of $\mathcal{K}$ to be. 

\section{Examples and applications}
\label{examples}

In this section we give two simple examples of misalignment with cosets $SU(3)/SO(3)$ and $SU(2)\times SU(2)/SU(2)$ and then move on to discuss some phenomenological applications in the context of composite Higgs models.

\subsection{$SU(3)/SO(3)$}
\label{SU3sect}
The coset $G/H=SU(3)/SO(3)$ is the simplest case where all the issues discussed above can be illustrated. This coset can be realized with hyperfermions $\Psi$ in any real irrep of a hypercolor group \GHC. Only few irreps can be used after requiring asymptotic freedom and the presence of a \GHC~singlet in $\Psi^3$. The adjoint irrep is always a possibility, although it is likely that these models fall into the conformal window. A safer option is to chose the anti-symmetric irrep of \GHC$=Sp(2N)$ with $N>2$, which has been already put on the lattice without hyperfermions~\cite{Bennett:2020qtj}.

In the vacuum $|{\mathrm{vac}}\rangle_0$ the generators of $SU(3)$ can be split into real symmetric broken generators 
$T^{\hat{1},\hat{2},\hat{3},\hat{4},\hat{5}} = \lambda^{1,3,4,6,8}$ and unbroken imaginary anti-symmetric $T^{1,2,3}=\lambda^{2,5,7}$, where $\lambda$ are the Gell-Mann matrices.
The mass matrix of $\Psi$, as shown in (\ref{canonicalmass}), can introduce explicit breaking, reducing the exact global symmetry group $S_{\mathrm{w}}$ to
\begin{align}
   S_{\mathrm{w}} = \left\{ \begin{array}{lll} 
   SU(3) & \mbox{ for }  & m_1=m_2=m_3=0 \\
    SO(3) & \mbox{ for }  & m_1=m_2=m_3\not=0  \\
    U(1) & \mbox{ for }  & m_1=m_2 \not= m_3  \\
    \{1\} & \mbox{ for }  & m_1\not =m_2 \not=m_3.
    \end{array} \right.
\end{align}
As discussed above, neither masses, nor gauging $SO(3)$ or $U(1)$, will misalign the vacuum.

One can misalign using only gauge coupling, by the method of~\cite{Dugan:1984hq}, gauging two commuting generators of $SU(3)$, i.e. $T^1$ and $T^{\hat{5}}$. (The group being gauged is necessarily partly chiral. Thus one must set the masses to zero and introduce additional spectator fermions to cancel the anomaly.) The squared mass matrix for the five pNGBs turn out to be proportional to 
\begin{align}
   M^2 \propto \mbox{diag}(4g^2, 4g^2, g^2-3\hat{g}^2, g^2-3\hat{g}^2, 0)  \label{twoU1s}
\end{align}
where $g$ and $\hat{g}$ are associated to the generators $T^1$ and $T^{\hat{5}}$ respectively. Varying the two gauge couplings, we see that as soon as $ g^2-3\hat{g}^2<0$ the two pNGBs associated with $T^{\hat{3}}$ and $T^{\hat{4}}$ become tachyonic, the vacuum $|{\mathrm{vac}}\rangle_0$ is misaligned, and both gauge bosons acquire a mass.

Alternatively, more relevant for this work, one can use fermionic couplings as in the previous section.
For definiteness, let us take $m_1=m_2=m_3\equiv m\not=0$, and gauge the $U(1)$ generator $T^1$, with coupling $g$.
The hyperquark mass contributes the same positive constant term to each of the five pNGBs, the gauge coupling contributes as in (\ref{twoU1s}) with $\hat{g}=0$, i.e. $M^2 \propto g^2 \mbox{diag}(4, 4, 1, 1, 0)$,
and neither of these contribution misaligns the vacuum.

Consider now a four-Fermi interaction involving a composite fermion operator in the symmetric ($\mathbf{6}$) of $SU(3)$ and an elementary fermion $\lambda$ charged under the gauged $U(1)$. The decomposition $SU(3) \to SO(3) \to U(1)$ is $\mathbf{6} \to \mathbf{1} + \mathbf{5} \to  (0_{\mathbf{1}}) + (0_{\mathbf{5}}) + (\pm 1_{\mathbf{5}}) + (\pm 2_{\mathbf{5}})$. Considering one elementary fermion characterized by the above $U(1)$ quantum number, the induced pNGB mass matrices are
\begin{align}
   M^2(0_{\mathbf{1}}) &\propto  y^2 {\mathrm{diag}}(2, 2, 2, 2, 2)  &M^2(0_{\mathbf{5}}) &\propto y^2 {\mathrm{diag}}(0,0,0,0, -2)\nonumber\\
   M^2(\pm 1_{\mathbf{5}}) &\propto y^2 {\mathrm{diag}}(0, 0, -1, -1, 0)  &M^2(\pm 2_{\mathbf{5}}) &\propto y^2 {\mathrm{diag}}(-1, -1, 0, 0, 0).
   \label{signsSU3SO3}
\end{align}

The overall sign of these contributions, given by the sign of $\mathcal{K}$ in (\ref{spectral}) is not  determined, although in cases like this one could speculate that the $H$-singlet should not misalign, thus the overall signs should be as in (\ref{signsSU3SO3}). One could further speculate a mass hierarchy $M_{\mathbf{5}} > M_{\mathbf{1}}$ between the composite fermions, assuming the dominance of the leading resonance in the spectral functions. 

If a gauge coupling is also present (as it must in phenomenologically relevant applications), one must also assume that the perturbatively irrelevant dimension-six operators are sufficiently enhanced in the infra-red in order to compete with the marginal gauge couplings. This is a crucial dynamical question about the hypercolor gauge theory, but may not be so far-fetched when the theory lies close to the strongly coupled edge of the conformal window. The same discussion applies to all subsequent cases.

\subsection{$SU(2)\times SU(2)/SU(2)$}
\label{SU2SU2sect}
This is just the case of two-flavors QCD, so the contribution of masses and gauging are well known. Here too, gauging a chiral group larger than $SU(2)$ misaligns the vacuum. In QCD, this is the well know statement that even in the absence of a Higgs field the $W$ and $Z$ bosons would acquire a small mass from the QCD quark condensate. This coset, however, can arise in other beyond the Standard Model scenarios addressing the dynamics of EWSB.
Consider now composite fermionic operators in the $({\mathbf{2}},{\mathbf{2}})$ and gauge a $U(1)$ subgroup (generated by $\sigma^3$) of the unbroken $SU(2)$. The decomposition $SU(2)\times SU(2) \to SU(2) \to U(1)$ is $({\mathbf{2}}, {\mathbf{2}})\to {\mathbf{1}} + {\mathbf{3}} \to  (0_{\mathbf{1}}) + (0_{\mathbf{3}}) + (\pm 1_{\mathbf{3}})$ yielding
\begin{align}
   M^2(0_{\mathbf{1}}) \propto  y^2 {\mathrm{diag}}(2, 2, 2)\quad  M^2(0_{\mathbf{3}}) \propto y^2 {\mathrm{diag}}(0,0,-2) \quad 
   M^2({\pm 1}_{\mathbf{3}}) \propto y^2 {\mathrm{diag}}(-1,-1,0).
\end{align}

Once again the overall sign of the pNGB masses can not be determined without an understanding of the strong dynamics. 
There are some estimates from the lattice on the spectrum of chimera baryons in a \GHC$=SU(4)$ gauge theory with four (Weyl) anti-symmetric $Q$ and two (Dirac) fundamental $q$ of $SU(4)$ \cite{Ayyar:2018zuk}. The hyperfermions $q$ lead to the $SU(2)\times SU(2)/SU(2)$ coset. The chimera baryons are $SU(4)$ singlet spin-1/2 states with mixed representations, denoted by $Qqq$ which can be decomposed into ${\mathbf{1}} + {\mathbf{3}}$ of the unbroken $SU(2)$. The results from  \cite{Ayyar:2018zuk} show that the masses of ${\mathbf{1}}$ and ${\mathbf{3}}$ are almost overlapping with each other within numerical uncertainties. In this case it is difficult to determine the overall sign of the pNGB masses using the lattice results.  

\subsection{Phenomenological applications}

The main application of vacuum misalignment is in the context of composite Higgs models. The three minimal cosets arising from 4D confining gauge theories that can generate a Higgs doublet and preserve custodial symmetry are $SU(4)/Sp(4)$ \cite{Barnard:2013zea,Ferretti:2013kya,Cacciapaglia:2014uja,Ferretti:2014qta,Agugliaro:2016clv,Galloway:2016fuo,Bennett:2022yfa}, $SU(5)/SO(5)$ \cite{Golterman:2015zwa,Ferretti:2014qta,Ferretti:2016upr,Golterman:2017vdj,Agugliaro:2018vsu} and $SU(4)\times SU(4)/SU(4)$ \cite{Ma:2015gra,Ferretti:2016upr}. In these models, the elementary top quark is assumed to couple with the composite top-partners through partial compositeness interactions. The projector matrices are chosen in such a way that the left handed quark doublet ($q_L$) is embedded in $(\mathbf{2},\mathbf{2})$  and the right handed top quark ($t_R$) in $(\mathbf{1},\mathbf{1})$ or $(\mathbf{1},\mathbf{3})$ under the unbroken $SU(2)_L\times SU(2)_R$ subgroup. Detailed constructions of these models can be found in \cite{Banerjee:2022izw}. 

Consider the coset $SU(4)/Sp(4)$ which gives rise to a Higgs doublet and a Standard Model gauge singlet $\eta$. A composite top-partner in the anti-symmetric $\mathbf{6}$ of $SU(4)$ decomposes as $\mathbf{6} \to \mathbf{1} + \mathbf{5}$ under $Sp(4)$.
The singlet $\mathbf{1}$ can couple to $t_R$ at leading order, while the $\mathbf{5} \to (\mathbf{2},\mathbf{2}) + (\mathbf{1},\mathbf{1})$ under $Sp(4) \to SU(2)_L\times SU(2)_R$ can couple to both $q_L$ or $t_R$.
As discussed earlier, we expect that the singlet $\mathbf{1}$ should not misalign the vacuum. Thus we expect a tachyonic mass term for the Higgs doublet to arise from the $(\mathbf{2},\mathbf{2})\in {\mathbf{5}}$ contribution, while the $(\mathbf{1},\mathbf{1})\in {\mathbf{5}}$ destabilizes the $\eta$ direction at LO.

The adjoint $\mathbf{15} \to \mathbf{5} + \mathbf{10}$ of $SU(4) \to Sp(4)$ can also misalign the vacuum. In contrast to the anti-symmetric case, here $q_L$ and $t_R$ can couple to both the $\mathbf{5}$ and $\mathbf{10}$ components. The $q_L$ generates mass terms for both the Higgs and $\eta$, while $t_R$ contributes only to the Higgs mass at LO. Without any input from the lattice, it is harder to identify if the misalignment is caused by the couplings of $\mathbf{10}$ or $\mathbf{5}$. However, for this particular scenario, recent preliminary results in the quenched approximation from \cite{Hsiao:2022kxf} provide more insight. The results in \cite{Hsiao:2022kxf} indicate that the chimera baryons transforming as $\mathbf{5}$ (denoted by $\Lambda$) could be heavier than the ones transforming as $\mathbf{10}$ (denoted by $\Sigma$) in a $Sp(4)$ hypercolor theory. Thus, the analysis presented in the previous section, together with this input from the lattice, suggests that the contribution from $\mathbf{5}$ is more likely to misalign the vacuum. 

Similar arguments can be offered for the $\mathbf{15}$ and $\mathbf{24}$ in the $SU(5)/SO(5)$, and for $(\mathbf{4},\mathbf{4})$ and $(\mathbf{4},\mathbf{\Bar{4}})$ in the $SU(4)\times SU(4)/SU(4)$ cosets.

\section{Conclusions and outlook}
\label{concl}

In this paper we shed some new light on the vacuum misalignment mechanism in strongly coupled gauge theories. In particular, we discussed the role of four-Fermi interactions between elementary fermions and composite fermionic operators in destabilizing the reference vacuum. We reviewed the well known results that hyperfermion masses or weak gauging of a subgroup of the unbroken global $H$ can not misalign the vacuum. Therefore, introducing the four-Fermi interactions is of particular importance for the models of dynamical EWSB in strongly coupled theory. We focused specifically on the partial compositeness couplings where an elementary fermion couples chirally to a composite spin-1/2 operator created by three hyperfermions. The potential generated at LO by these couplings can lead to a tachyonic mass term for the pNGBs, destabilizing the vacuum.

We showed that different representations of the composite operators contribute to the pNGB masses with different relative signs. Thus, irrespective of the overall sign of the effective Hamiltonian, there is always some representation that misaligns the vacuum.
We further pointed out that in case of a symmetric (anti-symmetric) irrep for the real (pseudoreal) coset, the expectation that the $H$-singlet operator does not misalign the vacuum can be used to predict the overall sign of the strong sector contributions. 
Lattice gauge theory calculations can help to test these expectations.

As a next step towards understanding the vacuum structure of the theory, one also needs to identify the directions on vacuum misalignment and the associated symmetry breaking pattern.
The main interest for EWSB lies in showing the existence of acceptable vacua only slightly misaligned from the unbroken one. This suggests that one should compute the full pNGB mass matrix $M^2$ in the unbroken vacuum \eqref{notachyon} (that depends on the hyperquark masses, the weak gauge couplings, and the partial compositeness couplings) and look for regions where one or more of its eigenvalues is on the verge of becoming tachyonic. This indicates which pNGB is going to acquire a vacuum expectation value once the eigenvalue becomes tachyonic, from which one can read off the pattern of spontaneous symmetry breaking. 

A generalization of the discussion presented in sections \ref{general} and \ref{partialcompo} to other composite operators, for instance with different spin, is possible. Instead of the partial compositeness one may also consider four-Fermi interaction between two elementary fermions and two hyperfermions ($\lambda\lambda\Psi\Psi$). In the low energy theory this would lead to a Yukawa type interaction among two elementary fermions and a composite scalar operator. The commutators between the charges and these scalar operators can be written in the same way as \eqref{QB_comm}. In this scenario the leading order Hamiltonian simply renormalizes the hyperfermion propagator. Therefore it can not misalign the vacuum. The next-to-leading order contributions can also be analyzed by extending the discussions presented in the section \ref{partialcompo}. 

\section*{Acknowledgments}

We would like to thank Maarten Golterman, Deog Ki Hong, Maurizio Piai, and Yigal Shamir for valuable comments on the manuscript. We also thank Ulf Gran for pointing out Cosmic cat to us. This work is supported by the Knut and Alice Wallenberg foundation under the grant KAW 2017.0100 (SHIFT project). 

\bibliography{reference.bib}

\providecommand{\href}[2]{#2}\begingroup\raggedright\begin{thebibliography}{10}

\bibitem{Dimopoulos:1979es}
S.~Dimopoulos and L.~Susskind, \emph{{Mass Without Scalars}},
  \href{http://dx.doi.org/10.1016/0550-3213(79)90364-X}{\emph{Nucl. Phys. B}
  {\bf 155} (1979) 237--252}.

\bibitem{Susskind:1978ms}
L.~Susskind, \emph{{Dynamics of Spontaneous Symmetry Breaking in the
  Weinberg-Salam Theory}},
  \href{http://dx.doi.org/10.1103/PhysRevD.20.2619}{\emph{Phys. Rev. D} {\bf
  20} (1979) 2619--2625}.

\bibitem{Kaplan:1983fs}
D.~B. Kaplan and H.~Georgi, \emph{{$SU(2) \times U(1)$ Breaking by Vacuum
  Misalignment}},
  \href{http://dx.doi.org/10.1016/0370-2693(84)91177-8}{\emph{Phys. Lett. B}
  {\bf 136} (1984) 183--186}.

\bibitem{Banks:1984gj}
T.~Banks, \emph{{Constraints on $SU(2) \times U(1)$ breaking by vacuum
  misalignment}},
  \href{http://dx.doi.org/10.1016/0550-3213(84)90389-4}{\emph{Nucl. Phys. B}
  {\bf 243} (1984) 125--130}.

\bibitem{Vafa:1983tf}
C.~Vafa and E.~Witten, \emph{{Restrictions on Symmetry Breaking in Vector-Like
  Gauge Theories}},
  \href{http://dx.doi.org/10.1016/0550-3213(84)90230-X}{\emph{Nucl. Phys. B}
  {\bf 234} (1984) 173--188}.

\bibitem{Witten:1983ut}
E.~Witten, \emph{{Some Inequalities Among Hadron Masses}},
  \href{http://dx.doi.org/10.1103/PhysRevLett.51.2351}{\emph{Phys. Rev. Lett.}
  {\bf 51} (1983) 2351}.

\bibitem{Weingarten:1983uj}
D.~Weingarten, \emph{{Mass Inequalities for QCD}},
  \href{http://dx.doi.org/10.1103/PhysRevLett.51.1830}{\emph{Phys. Rev. Lett.}
  {\bf 51} (1983) 1830}.

\bibitem{Dugan:1984hq}
M.~J. Dugan, H.~Georgi and D.~B. Kaplan, \emph{{Anatomy of a Composite Higgs
  Model}}, \href{http://dx.doi.org/10.1016/0550-3213(85)90221-4}{\emph{Nucl.
  Phys. B} {\bf 254} (1985) 299--326}.

\bibitem{ArkaniHamed:2002qy}
N.~Arkani-Hamed, A.~G. Cohen, E.~Katz and A.~E. Nelson, \emph{{The Littlest
  Higgs}}, \href{http://dx.doi.org/10.1088/1126-6708/2002/07/034}{\emph{JHEP}
  {\bf 07} (2002) 034}, [\href{https://arxiv.org/abs/hep-ph/0206021}{{\tt
  hep-ph/0206021}}].

\bibitem{Contino:2003ve}
R.~Contino, Y.~Nomura and A.~Pomarol, \emph{{Higgs as a holographic
  pseudoGoldstone boson}},
  \href{http://dx.doi.org/10.1016/j.nuclphysb.2003.08.027}{\emph{Nucl. Phys. B}
  {\bf 671} (2003) 148--174}, [\href{https://arxiv.org/abs/hep-ph/0306259}{{\tt
  hep-ph/0306259}}].

\bibitem{Agashe:2004rs}
K.~Agashe, R.~Contino and A.~Pomarol, \emph{{The Minimal composite Higgs
  model}}, \href{http://dx.doi.org/10.1016/j.nuclphysb.2005.04.035}{\emph{Nucl.
  Phys. B} {\bf 719} (2005) 165--187},
  [\href{https://arxiv.org/abs/hep-ph/0412089}{{\tt hep-ph/0412089}}].

\bibitem{Kaplan:1991dc}
D.~B. Kaplan, \emph{{Flavor at SSC energies: A New mechanism for dynamically
  generated fermion masses}},
  \href{http://dx.doi.org/10.1016/S0550-3213(05)80021-5}{\emph{Nucl. Phys. B}
  {\bf 365} (1991) 259--278}.

\bibitem{Barnard:2013zea}
J.~Barnard, T.~Gherghetta and T.~S. Ray, \emph{{UV descriptions of composite
  Higgs models without elementary scalars}},
  \href{http://dx.doi.org/10.1007/JHEP02(2014)002}{\emph{JHEP} {\bf 02} (2014)
  002}, [\href{https://arxiv.org/abs/1311.6562}{{\tt 1311.6562}}].

\bibitem{Ferretti:2013kya}
G.~Ferretti and D.~Karateev, \emph{{Fermionic UV completions of Composite Higgs
  models}}, \href{http://dx.doi.org/10.1007/JHEP03(2014)077}{\emph{JHEP} {\bf
  03} (2014) 077}, [\href{https://arxiv.org/abs/1312.5330}{{\tt 1312.5330}}].

\bibitem{Golterman:2015zwa}
M.~Golterman and Y.~Shamir, \emph{{Top quark induced effective potential in a
  composite Higgs model}},
  \href{http://dx.doi.org/10.1103/PhysRevD.91.094506}{\emph{Phys. Rev. D} {\bf
  91} (2015) 094506}, [\href{https://arxiv.org/abs/1502.00390}{{\tt
  1502.00390}}].

\bibitem{Ferretti:2014qta}
G.~Ferretti, \emph{{UV Completions of Partial Compositeness: The Case for a
  SU(4) Gauge Group}},
  \href{http://dx.doi.org/10.1007/JHEP06(2014)142}{\emph{JHEP} {\bf 06} (2014)
  142}, [\href{https://arxiv.org/abs/1404.7137}{{\tt 1404.7137}}].

\bibitem{DeGrand:2015yna}
T.~DeGrand and Y.~Shamir, \emph{{One-loop anomalous dimension of top-partner
  hyperbaryons in a family of composite Higgs models}},
  \href{http://dx.doi.org/10.1103/PhysRevD.92.075039}{\emph{Phys. Rev. D} {\bf
  92} (2015) 075039}, [\href{https://arxiv.org/abs/1508.02581}{{\tt
  1508.02581}}].

\bibitem{BuarqueFranzosi:2019eee}
D.~Buarque~Franzosi and G.~Ferretti, \emph{{Anomalous dimensions of potential
  top-partners}},
  \href{http://dx.doi.org/10.21468/SciPostPhys.7.3.027}{\emph{SciPost Phys.}
  {\bf 7} (2019) 027}, [\href{https://arxiv.org/abs/1905.08273}{{\tt
  1905.08273}}].

\bibitem{Ayyar:2018zuk}
V.~Ayyar, T.~Degrand, D.~C. Hackett, W.~I. Jay, E.~T. Neil, Y.~Shamir et~al.,
  \emph{{Baryon spectrum of SU(4) composite Higgs theory with two distinct
  fermion representations}},
  \href{http://dx.doi.org/10.1103/PhysRevD.97.114505}{\emph{Phys. Rev. D} {\bf
  97} (2018) 114505}, [\href{https://arxiv.org/abs/1801.05809}{{\tt
  1801.05809}}].

\bibitem{Hsiao:2022kxf}
H.~Hsiao, Bennett, D.~K. Hong, J.-W. Lee, C.~J.~D. Lin, B.~Lucini et~al.,
  \emph{{Spectroscopy of chimera baryons in a $Sp(4)$ lattice gauge theory}},
  \href{https://arxiv.org/abs/2211.03955}{{\tt 2211.03955}}.

\bibitem{Dashen:1970et}
R.~F. Dashen, \emph{{Some features of chiral symmetry breaking}},
  \href{http://dx.doi.org/10.1103/PhysRevD.3.1879}{\emph{Phys. Rev. D} {\bf 3}
  (1971) 1879--1889}.

\bibitem{Weinberg:1975gm}
S.~Weinberg, \emph{{Implications of Dynamical Symmetry Breaking}},
  \href{http://dx.doi.org/10.1103/PhysRevD.19.1277}{\emph{Phys. Rev. D} {\bf
  13} (1976) 974--996}.

\bibitem{Gell-Mann:1968hlm}
M.~Gell-Mann, R.~J. Oakes and B.~Renner, \emph{{Behavior of current divergences
  under $SU(3) \times SU(3)$}},
  \href{http://dx.doi.org/10.1103/PhysRev.175.2195}{\emph{Phys. Rev.} {\bf 175}
  (1968) 2195--2199}.

\bibitem{Das:1967it}
T.~Das, G.~S. Guralnik, V.~S. Mathur, F.~E. Low and J.~E. Young,
  \emph{{Electromagnetic mass difference of pions}},
  \href{http://dx.doi.org/10.1103/PhysRevLett.18.759}{\emph{Phys. Rev. Lett.}
  {\bf 18} (1967) 759--761}.

\bibitem{Peskin:1980gc}
M.~E. Peskin, \emph{{The Alignment of the Vacuum in Theories of Technicolor}},
  \href{http://dx.doi.org/10.1016/0550-3213(80)90051-6}{\emph{Nucl. Phys. B}
  {\bf 175} (1980) 197--233}.

\bibitem{Preskill:1980mz}
J.~Preskill, \emph{{Subgroup Alignment in Hypercolor Theories}},
  \href{http://dx.doi.org/10.1016/0550-3213(81)90265-0}{\emph{Nucl. Phys. B}
  {\bf 177} (1981) 21--59}.

\bibitem{Weinberg:1967kj}
S.~Weinberg, \emph{{Precise relations between the spectra of vector and axial
  vector mesons}},
  \href{http://dx.doi.org/10.1103/PhysRevLett.18.507}{\emph{Phys. Rev. Lett.}
  {\bf 18} (1967) 507--509}.

\bibitem{Golterman:2017vdj}
M.~Golterman and Y.~Shamir, \emph{{Effective potential in ultraviolet
  completions for composite Higgs models}},
  \href{http://dx.doi.org/10.1103/PhysRevD.97.095005}{\emph{Phys. Rev. D} {\bf
  97} (2018) 095005}, [\href{https://arxiv.org/abs/1707.06033}{{\tt
  1707.06033}}].

\bibitem{PhysRevLett.18.761}
T.~Das, V.~S. Mathur and S.~Okubo, \emph{Symmetry, superconvergence, and sum
  rules for spectral functions},
  \href{http://dx.doi.org/10.1103/PhysRevLett.18.761}{\emph{Phys. Rev. Lett.}
  {\bf 18} (May, 1967) 761--764}.

\bibitem{Bernard:1975cd}
C.~W. Bernard, A.~Duncan, J.~LoSecco and S.~Weinberg, \emph{{Exact Spectral
  Function Sum Rules}},
  \href{http://dx.doi.org/10.1103/PhysRevD.12.792}{\emph{Phys. Rev. D} {\bf 12}
  (1975) 792}.

\bibitem{Bennett:2020qtj}
E.~Bennett, J.~Holligan, D.~K. Hong, J.-W. Lee, C.~J.~D. Lin, B.~Lucini et~al.,
  \emph{{Glueballs and strings in $Sp(2N)$ Yang-Mills theories}},
  \href{http://dx.doi.org/10.1103/PhysRevD.103.054509}{\emph{Phys. Rev. D} {\bf
  103} (2021) 054509}, [\href{https://arxiv.org/abs/2010.15781}{{\tt
  2010.15781}}].

\bibitem{Cacciapaglia:2014uja}
G.~Cacciapaglia and F.~Sannino, \emph{{Fundamental Composite (Goldstone) Higgs
  Dynamics}}, \href{http://dx.doi.org/10.1007/JHEP04(2014)111}{\emph{JHEP} {\bf
  04} (2014) 111}, [\href{https://arxiv.org/abs/1402.0233}{{\tt 1402.0233}}].

\bibitem{Agugliaro:2016clv}
A.~Agugliaro, O.~Antipin, D.~Becciolini, S.~De~Curtis and M.~Redi, \emph{{UV
  complete composite Higgs models}},
  \href{http://dx.doi.org/10.1103/PhysRevD.95.035019}{\emph{Phys. Rev. D} {\bf
  95} (2017) 035019}, [\href{https://arxiv.org/abs/1609.07122}{{\tt
  1609.07122}}].

\bibitem{Galloway:2016fuo}
J.~Galloway, A.~L. Kagan and A.~Martin, \emph{{A UV complete partially
  composite-pNGB Higgs}},
  \href{http://dx.doi.org/10.1103/PhysRevD.95.035038}{\emph{Phys. Rev. D} {\bf
  95} (2017) 035038}, [\href{https://arxiv.org/abs/1609.05883}{{\tt
  1609.05883}}].

\bibitem{Bennett:2022yfa}
Bennett, D.~K. Hong, H.~Hsiao, J.-W. Lee, C.~J.~D. Lin, B.~Lucini et~al.,
  \emph{{Lattice studies of the Sp(4) gauge theory with two fundamental and
  three antisymmetric Dirac fermions}},
  \href{http://dx.doi.org/10.1103/PhysRevD.106.014501}{\emph{Phys. Rev. D} {\bf
  106} (2022) 014501}, [\href{https://arxiv.org/abs/2202.05516}{{\tt
  2202.05516}}].

\bibitem{Ferretti:2016upr}
G.~Ferretti, \emph{{Gauge theories of Partial Compositeness: Scenarios for
  Run-II of the LHC}},
  \href{http://dx.doi.org/10.1007/JHEP06(2016)107}{\emph{JHEP} {\bf 06} (2016)
  107}, [\href{https://arxiv.org/abs/1604.06467}{{\tt 1604.06467}}].

\bibitem{Agugliaro:2018vsu}
A.~Agugliaro, G.~Cacciapaglia, A.~Deandrea and S.~De~Curtis, \emph{{Vacuum
  misalignment and pattern of scalar masses in the SU(5)/SO(5) composite Higgs
  model}}, \href{http://dx.doi.org/10.1007/JHEP02(2019)089}{\emph{JHEP} {\bf
  02} (2019) 089}, [\href{https://arxiv.org/abs/1808.10175}{{\tt 1808.10175}}].

\bibitem{Ma:2015gra}
T.~Ma and G.~Cacciapaglia, \emph{{Fundamental Composite 2HDM: SU(N) with 4
  flavours}}, \href{http://dx.doi.org/10.1007/JHEP03(2016)211}{\emph{JHEP} {\bf
  03} (2016) 211}, [\href{https://arxiv.org/abs/1508.07014}{{\tt 1508.07014}}].

\bibitem{Banerjee:2022izw}
A.~Banerjee, D.~B. Franzosi and G.~Ferretti, \emph{{Modelling vector-like
  quarks in partial compositeness framework}},
  \href{http://dx.doi.org/10.1007/JHEP03(2022)200}{\emph{JHEP} {\bf 03} (2022)
  200}, [\href{https://arxiv.org/abs/2202.00037}{{\tt 2202.00037}}].

\end{thebibliography}\endgroup
\bibliographystyle{JHEP}
\end{document}